\documentstyle[12pt,epsf,epsfig]{article}
\include{graphicx}

\catcode`\@=11
\def\numberbysection{\@addtoreset{equation}{section}
\def\theequation{\thesection.\arabic{equation}}}
\numberbysection
\newcommand{\text}[1]{\textrm{#1}}
\def\stackunder#1#2{\mathrel{\mathop{#2}\limits_{#1}}}
\newcommand{\eq}{{\rm eq}}
\newcommand{\typ}{{\rm typ}}
\newcommand{\e}{{\rm e}}

\begin{document}

\author{J.M. Drouffe and C. Godr\`eche}
\title{Temporal correlations and persistence in the kinetic Ising model: 
the role
of temperature}
\maketitle

\begin{abstract}
We study the statistical properties of the sum
$S_t=\int_{0}^{t}dt'\,\sigma_{t'}$, 
that is the difference of time spent positive or negative
by the spin $\sigma_{t}$,
located at a given site of a $D$-dimensional Ising model 
evolving under Glauber dynamics from a random initial configuration.
We investigate the distribution of $S_{t}$ and
the first-passage statistics (persistence) of this
quantity.
We discuss successively the three regimes of high temperature
($T>T_{c}$), criticality ($T=T_c$), and low temperature ($T<T_{c}$).
We discuss in particular the question of the temperature dependence of the 
persistence exponent $\theta$, as well as that of the spectrum of exponents 
$\theta(x)$, in the low temperature phase.
The probability that the temporal mean $S_t/t$ was always larger
than the equilibrium magnetization is found to decay as $t^{-\theta-\frac12}$.
This yields a numerical determination of the
persistence exponent $\theta$ in the whole low temperature phase, 
in two dimensions, and above the roughening transition, 
in the low-temperature phase of the three-dimensional Ising model.

\end{abstract}

\section{Introduction}

Consider a $D$-dimensional lattice of Ising spins, evolving under Glauber
dynamics at a fixed temperature $T$ from a random initial configuration.
This is obtained e.g. by quenching the system from very high temperature
down to $T$. 
The purpose of this work is to study the influence of
temperature on the statistical properties of the sum 
\begin{equation}
S_{t}=\int_{0}^{t}dt^{\prime }\,\sigma _{t^{\prime }}, \label{sum}
\end{equation}
where $\sigma _{t}=\pm 1$ is the spin at a given site. More precisely, we
shall investigate two facets of this problem:

\begin{itemize}
\item the scaling of $S_{t}$ with $t$, and more generally the \emph{bulk}
properties of the distribution of $S_{t}$,

\item the statistics of rare events (such as large deviations, first
passages, persistence), i.e. the \emph{tail} properties of this
distribution.
\end{itemize}

Equivalent writings of the sum are 
$S_{t}=T_{t}^{+}-T_{t}^{-}=t\,M_{t}$, where 
\[
T_{t}^{\pm }=\int_{0}^{t}dt^{\prime }\,\frac{1\pm \sigma _{t^{\prime }}}{2} 
\]
are the lengths of time spent by the spin $\sigma _{t}$ in the positive
(negative) direction, or occupation times of the $\pm $ states, with $%
T_{t}^{+}+T_{t}^{-}=t$, and $M_{t}$ is the temporal mean of $\sigma _{t}$,
or local mean magnetization.
If one views $\sigma _{t}$ as the steps of a fictitious random walker, then 
$S_{t}$ is the position of the random walker at time $t$, and $M_{t}$ its
mean speed. 

Simple as it may seem, the problem thus stated is actually very intricate.
The reason is that, the values of $\sigma _{t}$ at
different instants of time being in general correlated, 
this study pertains to that of
sums of correlated random variables, for which no universal result exists.
The central limit theorem holds only for weakly correlated random variables.
As we shall see, $S_{t}$ obeys the central limit theorem for $T>T_{c}$,
while the latter is violated for $T\leq T_{c}$.

Hereafter, we focus our attention on the asymptotic behaviour of the
following quantities, as $t\rightarrow \infty $. \newline
\textbf{(i)} The probability distribution of $S_t$, 
or alternatively that of $M_t$,given by
\begin{equation}
P(t,x)={\mathcal{P}}\left( \frac{S_{t}}{t}>x\right) \qquad (-1<x<1),
\label{Ptx}
\end{equation}
with corresponding densities, related by
\[
f_{M}(t,x)=t\,f_{S}(t,y=tx)=-\frac{d}{dx}P(t,x).
\]
Assume that the typical value of $S_t$ scales, for long times, as $t^\alpha $.
By definition of $S_t$,  $\alpha\le 1$ necessarily.
Then, if $\alpha<1$, the events $\{S_{t}>tx\}$, with $x>0$, are rare.
(Or the events $\{S_{t}<tx\}$ with $x<0$.)
The tail probability $P(t,x)$, which measures their weight,
is vanishingly small, as $t\to\infty$.
In contrast, if $\alpha=1$, $P(t,x)$ is expected to converge to a limiting 
distribution,
i.e., the distribution of $S_t$ has fat tails.
The former case corresponds to $T\geq T_c$, the latter to $T<T_c$, as detailed below.

\noindent
\textbf{(ii)} The first-passage probability
\begin{equation}
R(t,x)={\mathcal{P}}\left( \frac{S_{t^{\prime }}}{t^{\prime }}>x,\text{ for
all }t^{\prime }\leq t\right) \qquad (-1<x<1), 
\label{Rtx}
\end{equation}
which is the probability for the random walker not to cross 
the line $S_{t'}=t'x$, up to time $t$.
This is also referred to as the probability of \emph{persistent large deviations}, 
sinceit involves the large-deviation event $\{S_{t}/t>x\}$, 
and the persistence condition $\{$for all $t^{\prime }\leq t\}$.

Special cases of these quantities are, first,
the probability of first passage of $S_{t}$ by the origin 
\begin{equation}
R(t,0)={\mathcal{P}}(S_{t^{\prime }}>0,\text{ for all }t^{\prime }\leq t),
\label{Rt0}
\end{equation}
and, secondly, the persistence probability of the process 
$\sigma_{t} $ (assuming that $\sigma _{t=0}=+1$)
\begin{equation}
p_{0}(t)={\mathcal{P}}(\text{no flip of }\sigma _{t^{\prime }},\text{ for all }%
t^{\prime }\leq t)\equiv {\mathcal{P}}\left( S_{t}=t\right) , 
\label{p0}
\end{equation}
which is also formally equal to $R(t,x=1)$.

The principal motivation behind such an investigation comes from the problem of 
\emph{phase persistence}, where, for a system undergoing phase ordering, the
question posed is: \emph{``What is the probability for a given point of
space to remain in the same phase as time passes?''}.

For the particular case of $T=0$, say for a spin model,
the question above is answered by the knowledge of $p_0(t)$,
which decays at long times as $t^{-\theta}$, thus defining
the persistence exponent $\theta$ \cite{dbg}.
However, as shown in \cite{dg}, consideration of the
more general events $\{S_{t}/t>x\}$,
beyond that of the most extreme event $\{S_{t}=t\}$
(the spin never flipped), allows both
a stationary definition of the persistence exponent $\theta$,
as the edge singularity of $\lim_{t\to\infty} P(t,x)$, for $x\to\pm1$,
and the introduction of a whole spectrum of exponents $\theta(x)$,
through the temporal decay of $R(t,x)$.
The interest of introducing such concepts is strengthened by the fact that,
for $0<T<T_c$, the same definition of the persistence exponent holds,
now for $x\to\pm m_{\eq}$ \cite{I} (see below).

In essence, the change in viewpoint when considering
$P(t,x)$ and $R(t,x)$ instead of $p_0(t)$ amounts
to shifting from the original question posed above 
to the more general one: \emph{``How long did the
system remain in a given phase?''}.
This means in particular searching the distribution of the length of time 
spent by the system inthe given phase, 
or occupation time of the phase, directly related to the
sum (\ref{sum}). 
The statistics of the occupation time provides
information on the ergodic nature of the process.
Further developments on this theme can be found in 
refs.~\cite{newman1} to \cite{newman3}.

This work is a sequel and a completion of \cite{I}. 
We investigate the behaviour of the distribution of $S_t$ and of
$R(t,x)$ in the three temperature regimes, $T>T_c$, $T=T_c$, and $T<T_c$,
and in particular revisit the question of the definition of the 
persistence exponent in the
low-temperature phase.
We will mainly consider here the two-dimensional case, 
and incidentally comment on the case $D=3$.

\section{Temperature regimes}

In its heat-bath formulation, Glauber dynamics consists in updating the spin 
$\sigma _{t}$, located at a given site, with the probability 
\begin{equation}
{\mathcal{P}}\left( \sigma _{t+dt}
=+1\right) =\frac{1}{2}\left( 1+\tanh \frac{h_{t}}{T}\right) ,
\label{heat} 
\end{equation}
where $h_{t}$ is the local field at the given site, equal to the sum of the
neighbouring spins. Under this dynamics spins thermalize in their local
environment.

\begin{figure}[htpb]
\centerline{
\includegraphics[angle=0,width=\linewidth]{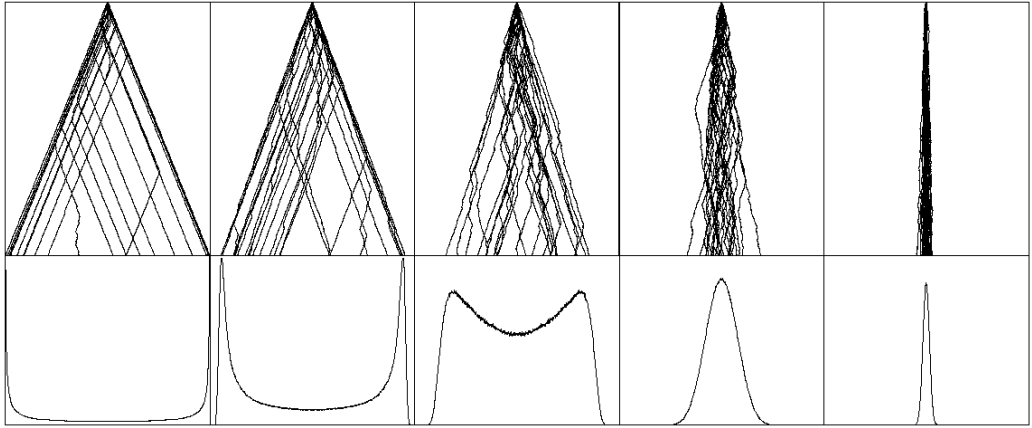}
}
\caption{Top curves: samples of spin histories.
From left to right: $T=0$, $T=0.9\,T_c$, $T=T_c$, 
$T=1.1\,T_c$, $T=\infty$.
The vertical axis is time ($0<t<1000$), the horizontal axis
represents $S_t$, the position of the Ising random walk.
Bottom curves: unnormalized histograms of $S_t$, at time $t=1000$.
(The system size is $4096^2$.)
}
\label{smp2}
\end{figure}

The behaviour of the system is qualitatively different according to the
value of the temperature at which the dynamics takes place, that is the
temperature $T$ after the quench. 
Three regimes are observed when letting $T$
vary from high ($T>T_{c}$) to low temperatures ($T<T_{c}$). 
An overview of the changes in behaviour with temperature is given 
in figure \ref{smp2}, which
depicts samples of spin histories, i.e., the position $S_{t}$ of the Ising
random walk as a function of time, for $0<t<1000$, together with the
corresponding distributions of $S_{t}$ at time $t=1000$, as temperature
varies.
The top pictures correspond to a sample of 30 spins located 
at various sites of a two-dimensional $4096^2$ square lattice,
the bottom pictures to all the spins of the lattice.
($T_c^{2D}=2/\ln (1+\sqrt{2})\approx 2.269$.)

\begin{figure}[htbp]
\centerline{
\includegraphics[angle=0,width=\linewidth]{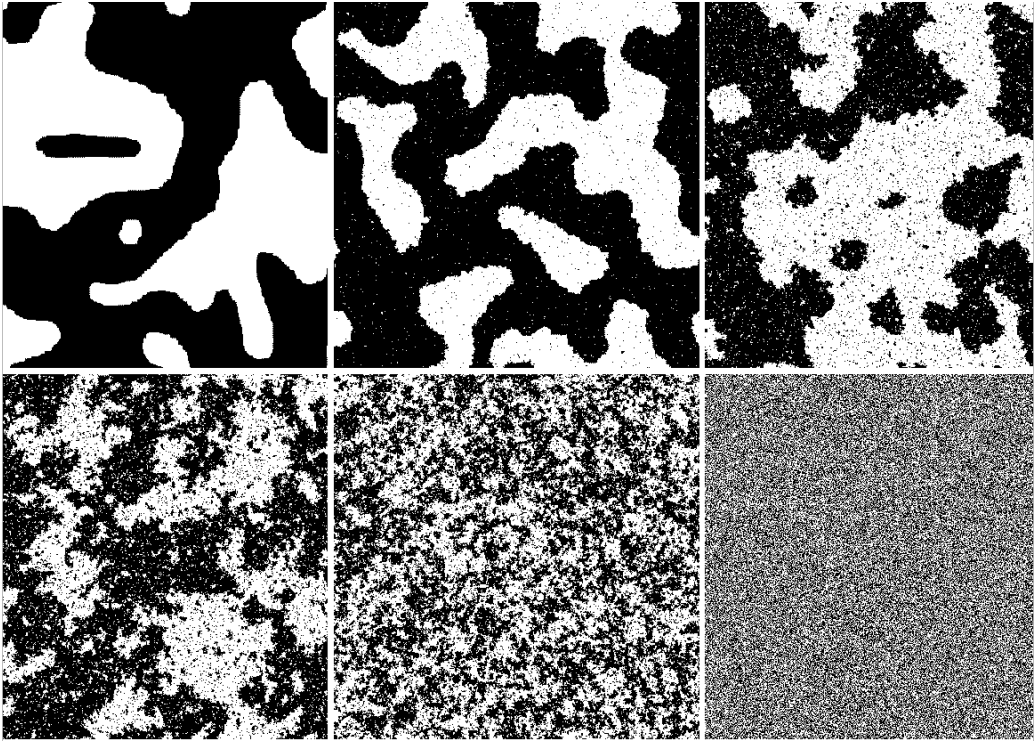}
}
\caption{Snapshots at time $t=1000$ of spin configurations 
for various temperatures.
From top left to bottom right: $T=0$, $0.7\,T_c$, $0.9\, T_c$, $T_c$, 
$1.1\, T_c$, and $T=\infty$.
}
\label{imgnb}
\end{figure}

\begin{figure}[htbp]
\centerline{
\includegraphics[angle=0,width=\linewidth]{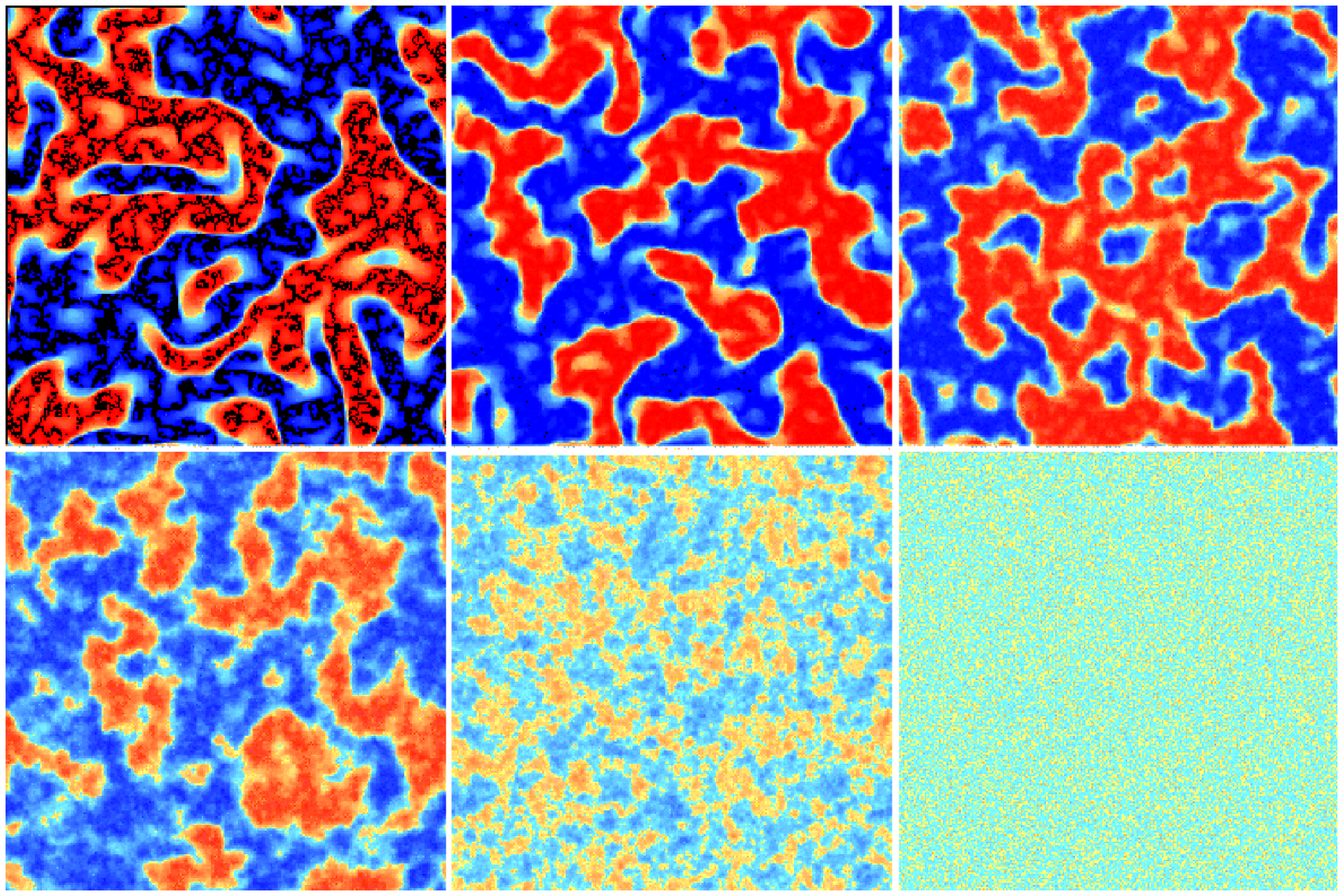}
}
\caption{Snapshots at time $t=1000$ of configurations 
of the values of $S_t$  for various temperatures.
From top left to bottom right: $T=0$, $0.7\,T_c$, $0.9\, T_c$, $T_c$, $1.1\, T_c$,
and $T=\infty$.
Values of $S_t$ ranging from $-1000$ to $1000$ are represented by
colours ranging from dark blue to dark red.
Persistent spins ($S_t=\pm 1000$), present in the top left picture ($T=0$) only,
are represented in black. }
\label{imgcoul}
\end{figure}

Snapshots at time $t=1000$ of spin configurations and of configurations 
of the values of $S_t$ for various temperatures are given in 
figures~\ref{imgnb} and \ref{imgcoul}.

\subsection{High temperature}

As long as $T>T_{c}$, thermal equilibrium is attained exponentially fast
(with a finite relaxation time $t_{\mathrm{eq}}$). 
The equilibrium
magnetization $m_{\mathrm{eq}}$ of the system is equal to $0$.

\begin{figure}[htbp]
\centerline{
\includegraphics[width=.75\linewidth]{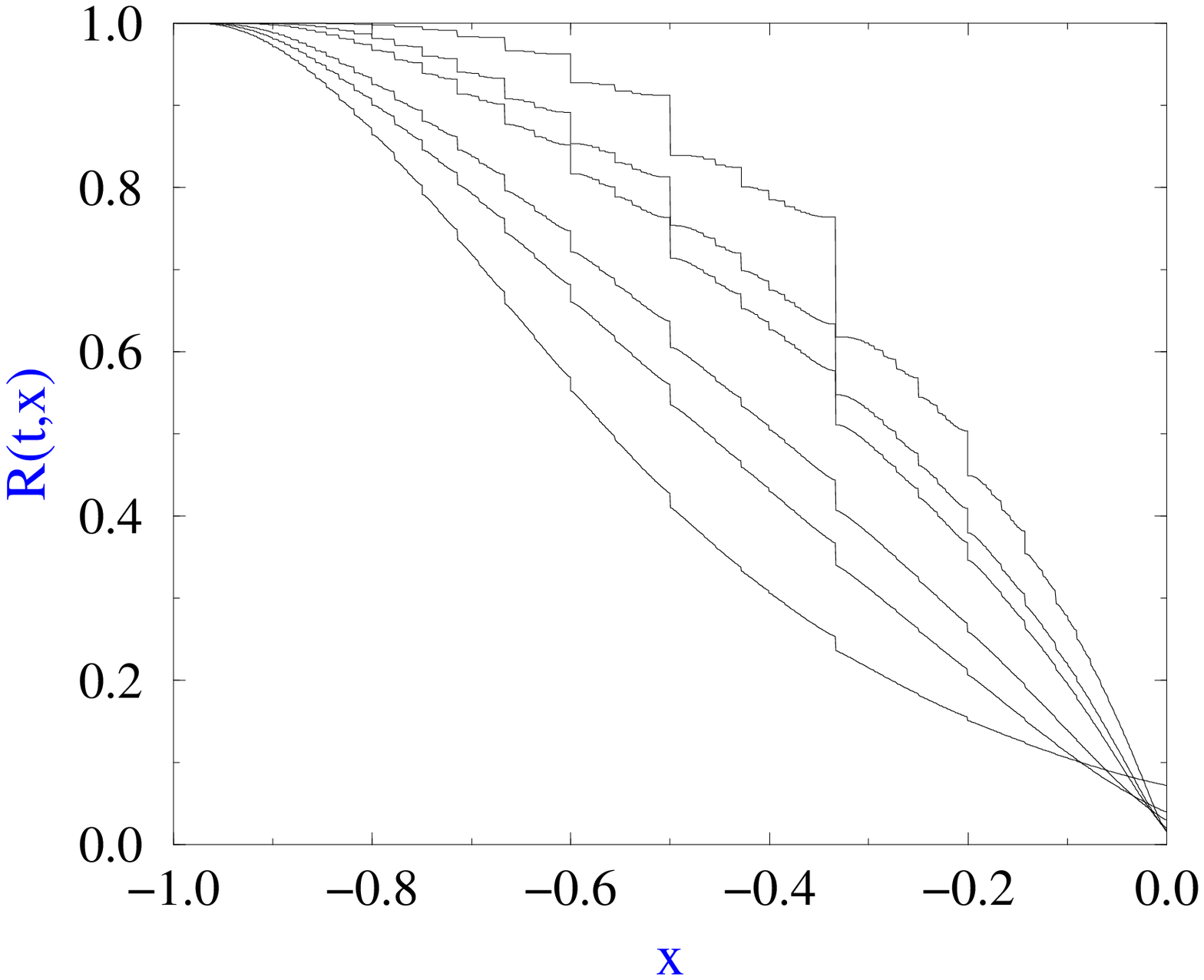}
}
\caption{$R(t=10000,x)$ in the high temperature phase, for $-1<x<0$. 
From top to bottom: $T=\infty$, $1.5\,T_c$, $1.3\,T_c$, $1.1\,T_c$, 
$1.05\,T_c$, $1.01\,T_c$. 
(The system size is $4096^2$.)
}
\label{fig_devil}
\end{figure}

Let us first consider the simplest case $T=\infty $. 
The values of the spin $\sigma _{t}$ at successive instants of time 
are independent and equiprobable, as the steps of a binomial random walk 
in continuous time,
hence $S_{t}$ is a sum of independent, identically distributed random
variables. 
Therefore the central limit theorem holds. 
The typical value of $S_{t}$ reads 
\[
(S_{t})_{\mathrm{typ}}\sim t^{\frac{1}{2}},
\]
and the limiting distribution of $S_{t}/t^{1/2}$ is Gaussian. 
More precisely, the bulk of the distribution reads 
\[
P(t,x=t^{-\frac{1}{2}}z)={\mathcal{P}}\left( t^{-\frac{1}{2}}S_{t}>z\right)
\sim \int_{z}^{\infty }du\,\exp \left( -\frac{u^{2}}{2}\right) ,
\]
because the variance of $S_{t}$ equals $t$. 
Hence, for (the bulk of) the density of $M_{t}$, we have 
\begin{equation}
f_{M}(t,x)\sim t^{\frac{1}{2}}\exp \left( -\frac{t\,x^{2}}{2}\right) .
\label{fM_high}
\end{equation}
Figure~\ref{smp2} gives an illustration of the random walk associated to $S_{t}$,
and of the Gaussian distribution of this quantity at $T=\infty$ 
(first pictures from the right).

If $t$ is large, and $x>0$, $P(t,x)$ measures the probability of
the rare events such that $S_{t}\sim t$,
or $M_{t}\sim O(1)$, i.e., of large deviations of $S_{t}$ with respect to its
typical behaviour. 
For any sum $S_{t}$ of independent, identically
distributed random variables, the tail of $P(t,x)$ (probability of a large
deviation) is given by 
\begin{equation}
P(t,x)\stackunder{t\to\infty}{\sim }\exp \left( -I(x)\,t\right) 
\qquad (x>0). 
\label{ptx_high}
\end{equation}
In the present case, $T=\infty $, that is for a binomial random walk (with
equiprobable steps $\pm 1$), we have $I(x)=\frac{1}{2}[ (1+x)\ln (1+x)$ 
$+(1-x)\ln (1-x)] $.
(For a simple derivation see \cite{dg}.)
For $x\rightarrow 0$, 
$I(x)\approx x^{2}/2$, the scaling variable is $x^{2}t=z^{2} $, hence the
central limit theorem is recovered.

\begin{figure}[htbp]
\centerline{
\includegraphics[width=.5\linewidth]{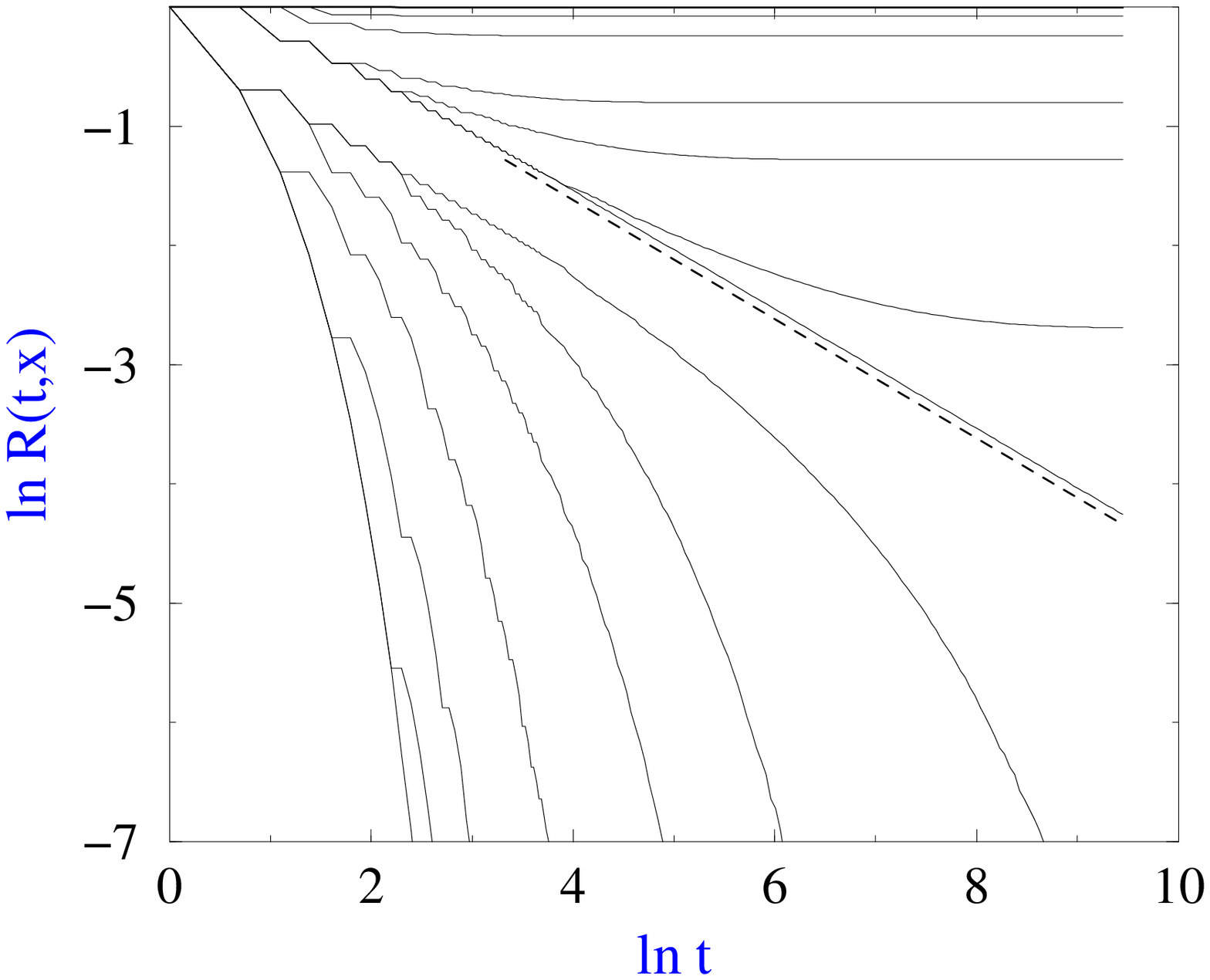}
\includegraphics[width=.5\linewidth]{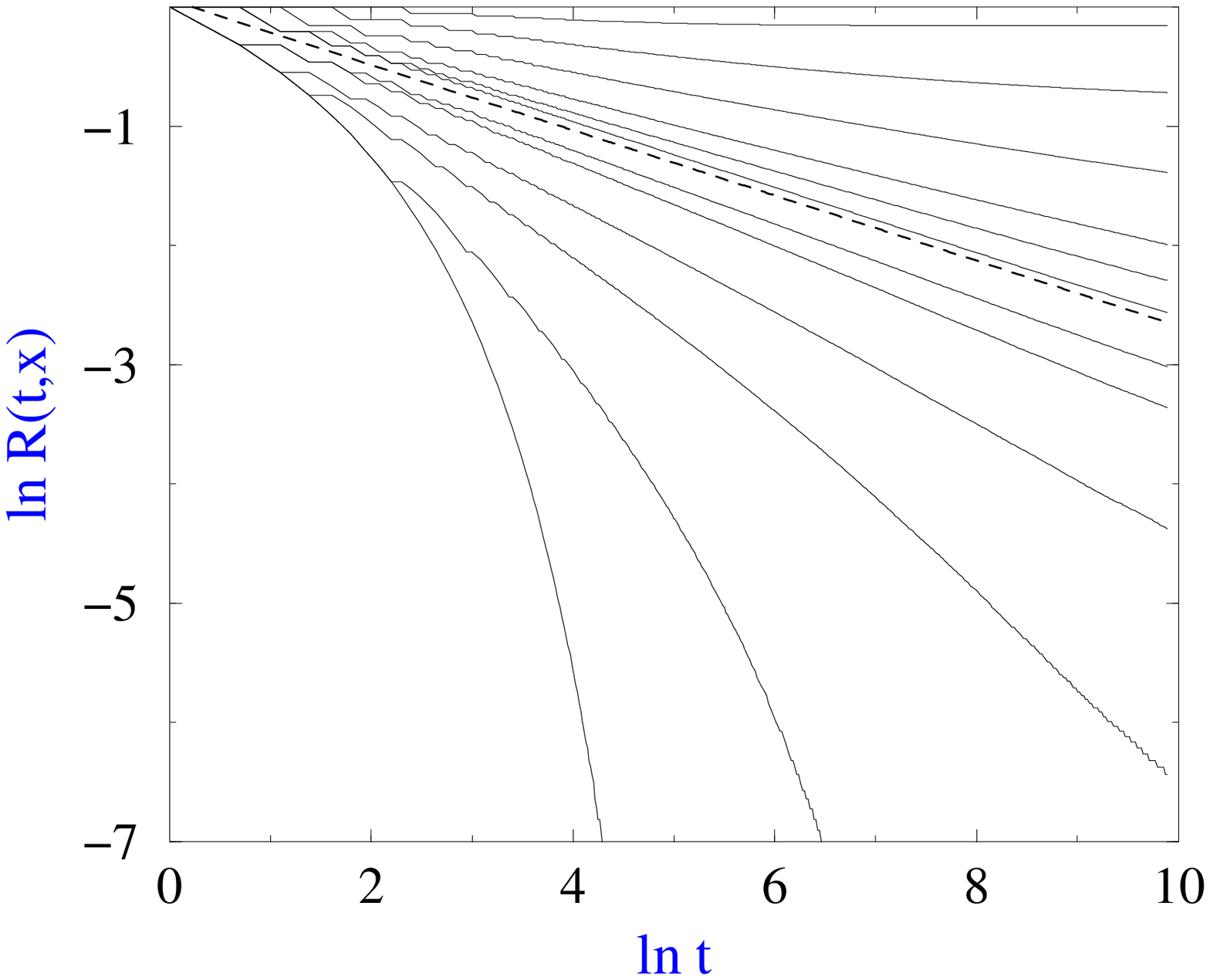}
}
\caption{$R(t,x)$ at $T=\infty$ (left) and $T=T_c$ (right). \newline
For $T=\infty$, from bottom to top: $x=$ 1, 0.8, 0.6, 0.4, 0.2, 0.1, 0.02, 0, 
$-0.02$, $-0.1$, $-0.2$, $-0.4$, $-0.6$, $-0.8$.
The dashed line parallel to $R(t,0)$ has slope $-\theta(0)=-\frac{1}{2}$.\newline
For $T=T_c$, from bottom to top: $x=$1, 0.8, 0.6, 0.4, 0.2, 0.1, 0,
$-0.1$, $-0.2$, $-0.4$, $-0.6$, $-0.8$.
The dashed line parallel to $R(t,0)$ has slope $-\theta(0)=-0.273$.
(The system size is $4096^2$.)}
\label{R_ht_cg}
\end{figure}

For a sum $S_{t}$ of independent random variables, the probability of first
passage by the origin, ${\mathcal{P}}(S_{t^{\prime }}>0,$ for all 
$t^{\prime}\leq t)$, decays as $t^{-\frac{1}{2}}$, as is well known. 
The behaviour of $R(t,x)$, for $x\neq0$, is more subtle. 
According to ref. \cite{bauer} we have, for a
binomial random walk, as $t\rightarrow \infty $, 
\begin{equation}
R(t,x)\sim \left\{ 
\begin{array}{l}
\exp \left( -I(x)t\right) \\ 
t^{-\frac{1}{2}} \\ 
R_{\infty }(x)
\end{array}
\begin{array}{l}
0<x<1 \\ 
x=0 \\ 
-1<x<0
\end{array}
\right. . 
\label{rtx_high}
\end{equation}
The asymptotic value $R_{\infty }(x)$ is the top curve of figure \ref{fig_devil}. 
It is
obtained by taking the value at time $t=10000$ of $R(t,x)$ for $T=\infty $,
depicted in figure~\ref{R_ht_cg} (left). 
The devil's staircase thus obtained, a
discontinuous curve at all the rationals, is indistinguishable from its
analytical prediction given in ref. \cite{bauer}.

The persistence probability $p_0(t)$ behaves, as $t\rightarrow \infty $,
as $p_{0}(t)\sim \mathrm{e}^{-(\ln 2)\,t}$, which matches both (\ref{ptx_high})
and (\ref{rtx_high}) for $x=1$. Note that (\ref{rtx_high}) defines the
first-passage exponent $\theta (x=0)=\frac{1}{2}$.

Let us now investigate the deformations induced on the quantities considered
above, when $T$ decreases.

If $T_{c}<T<\infty $, the two-time autocorrelation $\left\langle \sigma
_{s}\sigma _{t}\right\rangle $ (with $s<t$) is short ranged. At equilibrium,
i.e., for $1\ll s$, $\left\langle \sigma _{s}\sigma _{t}\right\rangle \sim
\exp \left( -(t-s)/t_{\mathrm{eq}}\right) $. 
Hence the statistics of $S_{t}$
is that of a random walk with short-ranged correlations, and the central
limit theorem still holds. 
The distribution of $S_{t}$ is Gaussian (see second
pictures from the right in figure~1, for $T=1.1\,T_{c}$), with a width
proportional to $(t\,t_{\eq})^\frac12$. 
While the Gaussian shape of $P(t,x)$ in
the bulk is universal, as long as the correlations $\left\langle \sigma
_{s}\sigma _{t}\right\rangle $ are short-ranged, the tails are not. 
Hence $I(x)$ is expected to be deformed as $T$ decreases from infinity to $T_{c}$.

The probability of first passage by the origin, ${\mathcal{P}}(S_{t^{\prime
}}>0,$ for all $t^{\prime }\leq t)$, still decays as $t^{-\frac{1}{2}}$. 
The persistence probability $R(t,x)$ behaves qualitatively as 
in equation~(\ref{rtx_high}).
As $T$ decreases down to $T_{c}$, the discontinuous curve $R_{\infty }(x)$
is deformed, as seen in figure~\ref{fig_devil}. 
Gaps vanish when $T\to T_c$.
Note the finite time effect when
approaching $T_{c}$, manifested by the fact that $R_{\infty }(x)$ does not
vanish as $x\rightarrow 0$. 

\subsection{Critical coarsening}

The system is now quenched from a disordered initial state to its critical
point.

After the quench, spatial correlations develop in the system, just as in the
critical state, but only over a length scale which grows like $t^{1/z_{c}}$,
where $z_{c}$ is the dynamic critical exponent. The equal-time correlation
function has the scaling form 
\[
C_{\mathbf{x}}(t)
=\left\langle \sigma _{t}(0)\sigma _{t}({\mathbf{x}})\right\rangle 
=|{\mathbf{x}}|^{-2\beta /\nu }\,
\phi \!\left( \frac{|{\mathbf{x}}|}{t^{1/z_{c}}}\right) ,
\]
where $\mathbf{x}$ is a lattice point, and $\beta $ and $\nu $ are the usual
static critical exponents. 
The scaling function $\phi (y)$ goes to a
constant for $y\to 0$, while it falls off exponentially to zero for $x\to
\infty $, i.e., on scales smaller than $t^{1/z_{c}}$ the system looks
critical, while on larger scales it is disordered. 
In two dimensions $z_{c}\approx 2.17$, $\beta =\frac{1}{8}$, 
and $\nu =1$; 
in three dimensions 
$z_{c}\approx 2.04$, $\beta\approx 0.327$, and $\nu\approx 0.63$ \cite{jaster}.

\begin{figure}[htbp]
\centerline{
\includegraphics[width=0.75\linewidth]{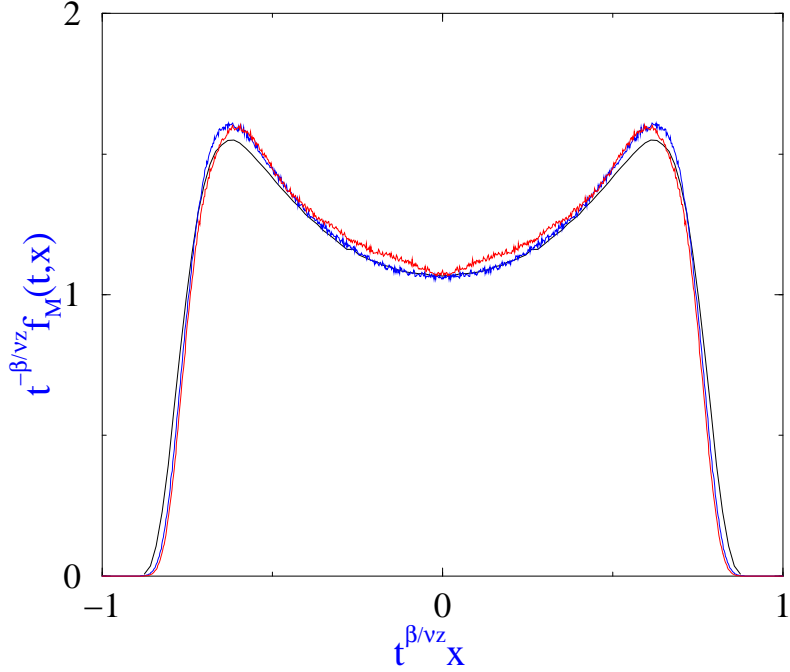}
}
\caption{Critical scaling function $\varphi$ obtained by rescaling $f_M(t,x)$
for three different times: $t= $100, 1000, 10000, see equation (\ref{fM_crit}).
(The system size is $4096^2$.)
}
\label{rescal_crit}
\end{figure}

For $1\ll s\sim t$ the two-time autocorrelation function scales 
as \cite{janssen,glFDT} 
\begin{equation}
C(t,s)=\left\langle \sigma _{s}\sigma _{t}\right\rangle =s^{-2\beta /\nu
z_{c}}\,g_{c}\left( \frac{s}{t}\right) . \label{cts_crit}
\end{equation}
A numerical determination of the scaling function $g_c$ can be found in 
ref.~\cite{glFDT}.
This implies, setting $a=2\beta /\nu z_{c}$, 
\begin{eqnarray*}
\left\langle M_{t}^{2}\right\rangle &=&\frac{2}{t^{2}}\int_{0}^{t}dt_{2}%
\int_{0}^{t_{2}}dt_{1}\,\left\langle \sigma _{t_{1}}\sigma
_{t_{2}}\right\rangle \\
&=&2t^{-a}\,\int_{0}^{1}dv\int_{0}^{v}du\,u^{-a}\,g_{c}\left( \frac{u}{v}%
\right) \\
&=&t^{-a}\left( \frac{2}{2-a}\int_{0}^{1}dx\,x^{-a}g_{c}(x)\right) ,
\end{eqnarray*}
hence 
\begin{equation}
(S_{t})_{\mathrm{typ}}\sim t^{1-\beta /\nu z_{c}}. 
\label{scal_crit}
\end{equation}
Note that this scaling behaviour was obtained from the \emph{non-stationary} form 
(\ref{cts_crit}).
The behaviour of $S_{t}$ at criticality is illustrated by the middle
pictures of figure~1. 

From equation (\ref{scal_crit}) one can infer that the bulk of
the probability density of $M_{t}$ scales as
\begin{equation}
f_{M}(t,x)\sim t^{\beta /\nu z_{c}}\varphi (t^{\beta /\nu z_{c}}\,x),
\label{fM_crit}
\end{equation}
which is the critical counterpart of (\ref{fM_high}). 
The scaling function $\varphi $ is depicted in figure~\ref{rescal_crit},
 for $D=2$, with $%
\beta /\nu z_{c}\approx 0.0576$. 
Due to the smallness of this exponent, $f_{M}$ is very slowly peaking. 
The existence of the scaling form (\ref
{fM_crit}) implies gap scaling for higher moments, i.e., 
$\left\langle
M_{t}^{2k}\right\rangle \sim \left\langle M_{t}^{2}\right\rangle ^{k}$. 

The probability $R(t,x)$ behaves qualitatively as in the high
temperature phase (see figure~\ref{R_ht_cg}, right). 
For $0<x<1$, it decreases to zero faster than a power-law
(in particular $p_{0}(t)\equiv R(t,x=1)$).
For $-1<x<0$,
it decreases, extremely slowly, to a constant $R_{\infty}(x)$ 
obtained by extrapolating the results of figure~\ref{fig_devil}, 
as $T\rightarrow T_{c}$.
Finally for $x=0$, $R(t,x)\sim t^{-\theta (0)}$. 
We find $\theta (0)\approx 0.27$ in
2D and $\theta (0)\approx 0.41$ in 3D. 
Note that $\theta (0)$ is a nonequilibrium {\it critical} exponent.

\textit{Remark}. 
The exponent $\theta (0)$ can also be measured at
criticality for the two-dimensional $p_{1}-p_{2}$ model \cite{drg,oliv}. 
This model is defined as follows.
The right side of (\ref{heat}), 
$p(h)=\frac12(1+\tanh h/T)$,
is a function of $h$ ($h=\pm4,\pm2,0$),
such that $p(-h)=1-p(h)$.
In particular $p(0)=\frac12$.
The dynamics of the Ising model therefore depends on one parameter only, 
which is $p_1=p(2)$, or alternatively $p_2=p(4)$, the two being related.
Considering instead these two quantities as independent 
($\frac12\le p_1,p_2\le 1$)
defines the $p_{1}-p_{2}$ model.
For the voter model ($p_{1}=\frac34$, $p_{2}=1$) one finds 
$\theta (0)\approx 0.37$. 
As one moves along the critical line, from thevoter point 
to the Ising critical point ($p_1=0.854$, $p_2=0.971$), 
numerical measurements show algebraic decay of $R(t,0)$,
with a seemingly constant slope on a log-log plot
(i.e. no sign of a crossover to $\theta (0)\approx 0.27$),
defining a decreasingexponent $\theta (0)$. 
Then, from the Ising critical point to the end of the
curve ($p_{1}=1$, $p_{2}\approx 0.85$), this exponent is constant,
$\theta (0)\approx 0.27$.

\subsection{Low temperature coarsening}

In the low-temperature phase $(0\leq T<$ $T_{c})$, the lower the temperature,
the higher the tendency of a spin to align with the majority.
Hence the system coarsens, i.e., domains of opposite signs grow,
because the system tries to reach locally one of the two equilibrium phases,
corresponding to an equilibrium magnetization $\pm m_{\mathrm{eq}}$, 
where (in 2D) 
\[
m_{\mathrm{eq}}(T)
=\left( 1-\left( \sinh 2/T\right) ^{-4}\right) ^{\frac{1}{8}}.
\]
In the scaling regime the system is statistically self-similar, with only
one single characteristic length scale, which is the size of a typical
domain.

For a typical spin, deep into a domain, two scales of time 
are observed: a fast one, due to thermal
flips, and a slow one, due to the passage of domain walls. 
The net result of the thermal flips can be seen on figure~1.
For example, for $T=0.9\,T_c$ (second pictures from the left), 
the distribution of $S_t$ concentrates, i.e., no longer covers the whole
interval $[-t,t]$.
A more precise statement is given presently.

In the regime $1\ll s\sim t$, the autocorrelation function scales as 
\begin{equation}
C(t,s)= m_{\mathrm{eq}}^{2}\,(T)\,g\left( \frac{s}{t}\right) .
\label{hyp}
\end{equation}
According to this hypothesis \cite{bray}, all the temperature dependence
is factored out in the prefactor $m_{\mathrm{eq}}^{2}$.
Thus, by the same computation as above, 
\begin{equation}
\left\langle M_{t}^{2}\right\rangle = m_{\mathrm{eq}%
}^{2}(T)\,\int_{0}^{1}dx\,g(x).
\label{scal}
\end{equation}
Hence 
\[
(S_{t})_{\mathrm{typ}}\sim t,
\]
which indicates the existence of a limiting distribution for $M_t$, 
as $t\to\infty$:
$$
f_M(x)=\lim_{t\to\infty}f_M(t,x).
$$
This distribution is observed to be
a U-shaped curve defined on $[-m_{\rm eq},m_{\rm eq}]$ \cite{I}.

Equation~(\ref{scal}) shows that the variance of 
$M_t/m_{\mathrm{eq}}$ no longer depends on temperature. 
Assuming that the same factorization of $m_{\eq}$ holds for higher moments implies
that the whole limiting distribution of the
rescaled variable $M_t/m_{\mathrm{eq}}$ is 
independent of temperature, and therefore identical to the $T=0$ distribution.
The latter being singular at $x\rightarrow \pm 1$, 
with singularity exponent equal to $\theta -1$ \cite{dg,I},
the existence of a unique master curve for the distribution of 
$M/m_{\mathrm{eq}}$
thus provides a simple definition of the persistence exponent
$\theta $ at finite temperature.

\begin{figure}[htbp]
\centerline{
\includegraphics[width=0.75\linewidth]{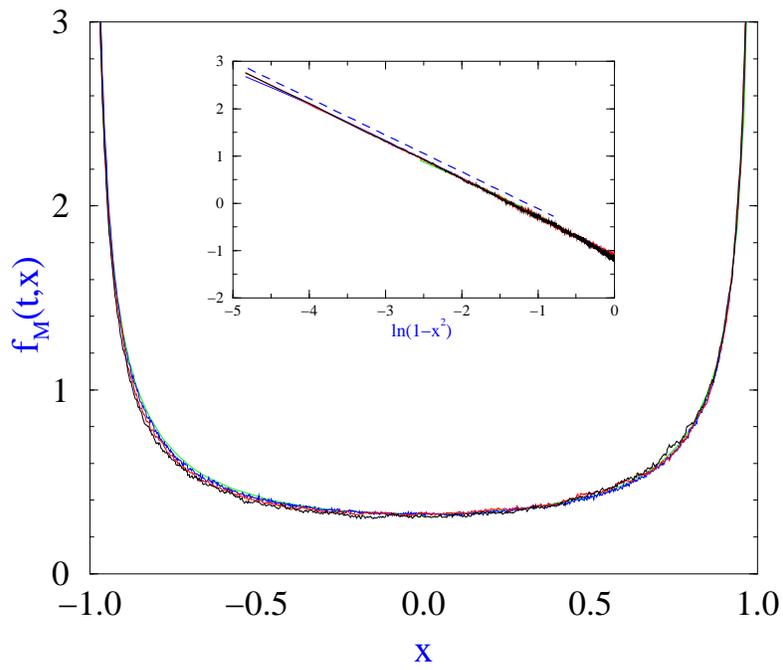}
}
\caption{Probability density $f_M(t,x)$ at $T=0$ for
$t=100$, 1000, 10000, 20000. 
Inset: log-log plot of the same curves against 
$1-x^2$.
The dashed line has slope $-0.78$.
(The system size is $4096^2$.)}
\label{dist0_cg}
\end{figure}

As seen in figure \ref{dist0_cg}, 
the zero-temperature limiting distribution $f_M$ is
extremely close to a beta distribution $\sim(1-x^2)^{\theta-1}$, hence,
in the low-temperature phase, we have
$$
f_M(x)\approx 
\frac{1}{m_{\eq}}f_M^{\rm beta}\left(\frac{x}{m_\eq}\right)
=\frac{1}{m_{\eq}}\frac{\Gamma(\frac{1}{2}+\theta )}{\sqrt{\pi }\Gamma (\theta ) }
\left(1-\frac{x^2}{m_{\eq}^2}\right)^{\theta-1}.
$$
In two dimensions, the determination of $\theta $ from $f_M$ at $T=0$ yields
$\theta \approx 0.22$ (see figure~\ref{dist0_cg}), 
confirming previous estimates obtained either by numerical simulations 
\cite{dbg,stauffer1}, or 
in an experiment on a liquid crystal system \cite{yurke}. 
(See also \cite{sire}.)

\begin{figure}[htbp]
\centerline{
\includegraphics[width=.5\linewidth]{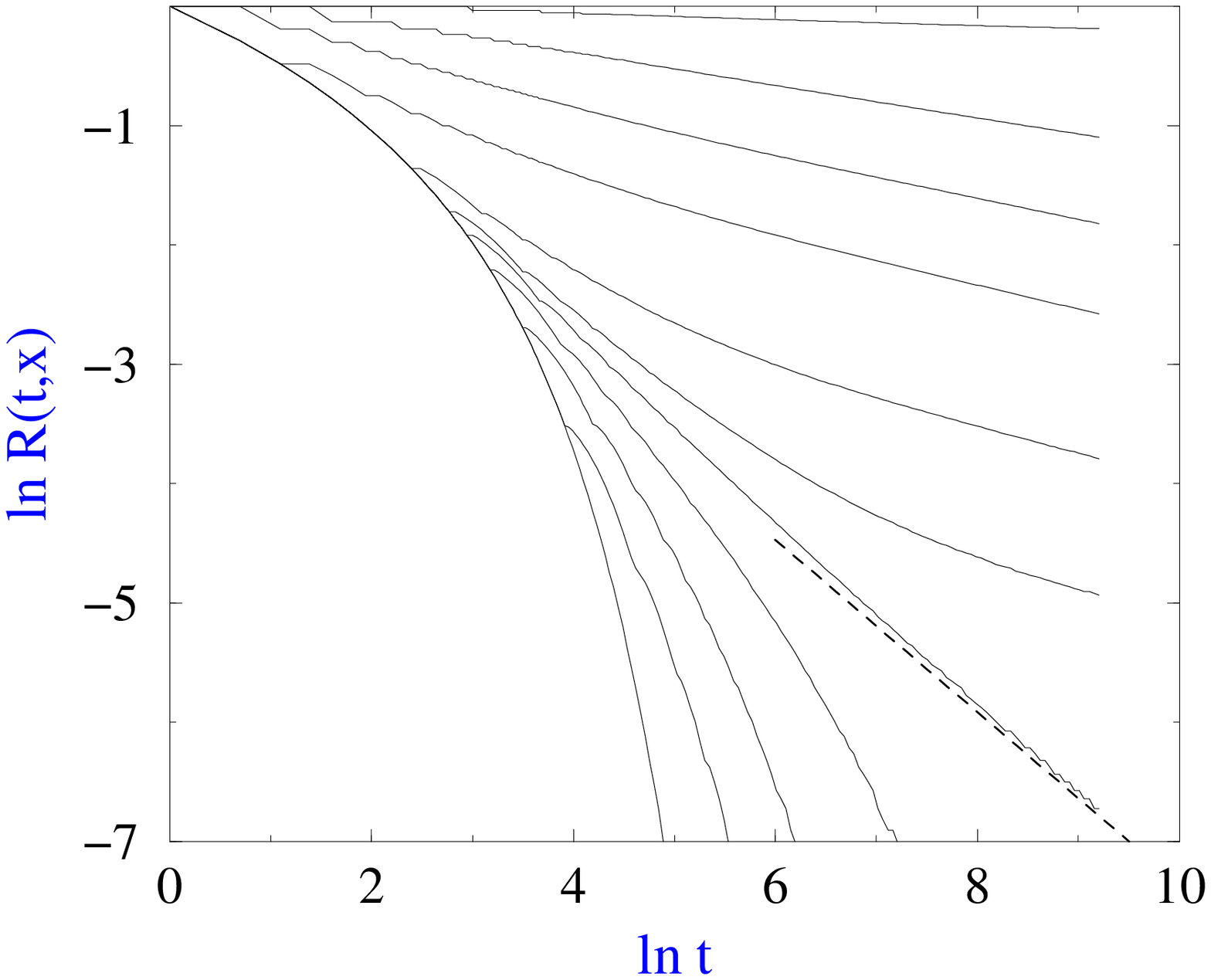}
\includegraphics[width=.5\linewidth]{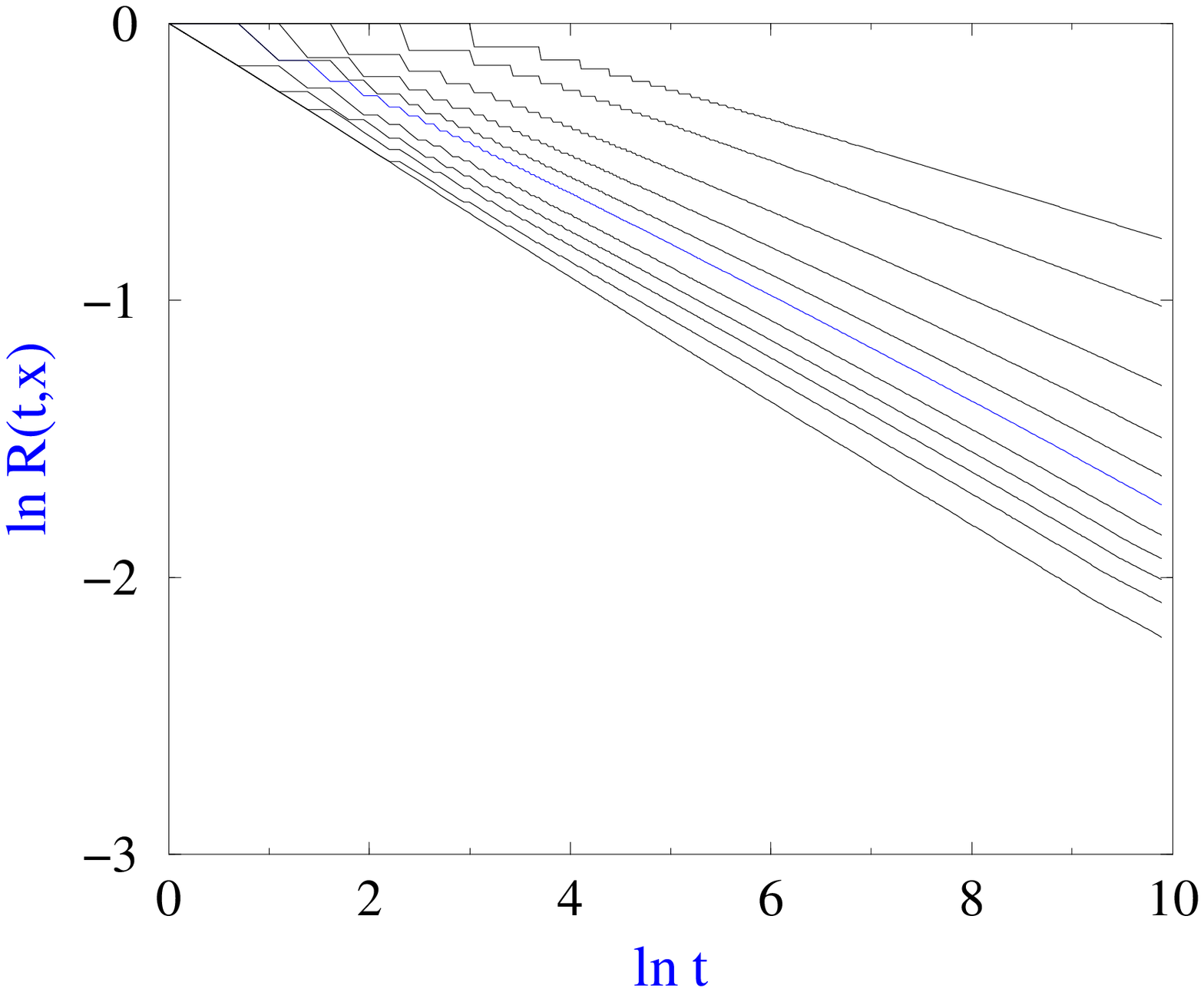}
}
\caption{$R(t,x)$ at $T=0.895\,T_c$ (corresponding to $m_{\rm{eq}}=0.9$)
(left) and $T=0$ (right).\newline
For $T=0.895\,T_c$, from bottom to top: 
$x=$ 1, 0.96, 0.94, 0.92, 0.9, 0.88, 0.82, 0.5, 0,
$-0.6$, $-0.9$.
The dashed line has slope $-\theta(m_{\rm eq})=-0.22-\frac{1}{2}$.\newline
For $T=0$, from bottom to top: $x=$ 1, 0.8, 0.6, 0.4, 0.2, 0,
$-0.2$, $-0.4$, $-0.6$, $-0.8$, $-0.9$.
(The system size is $4096^2$.)}
\label{r2dlt09_cg}
\end{figure}

Parallel observations come from the measurement of $R(t,x)$.
For $-m_{\mathrm{eq}}<x<m_{\mathrm{eq}}$, we have
\[
R(t,x)={\mathcal{P}}(S_{t^{\prime }}/t^{\prime }>x\text{ for all }t^{\prime
}\leq t)\sim t^{-\theta (x)}, 
\]
which defines the spectrum of exponents $\theta (x)$ \cite{dg,I}.
Outside this interval, $R(t,x)$ either decays faster than a power-law, 
or goes to a constant (see figure~\ref{r2dlt09_cg}).
In two dimensions we observed with reasonable accuracy
that $\theta (x/m_{\mathrm{eq}})$ does not depend on temperature.
This generalizes the observation, made in \cite{I}, 
for the two-dimensional Ising model,
that the first-passage exponent $\theta(0)$ ($\approx 0.19$ in 2D)
is independent of temperature for any $T<T_c$.

\begin{figure}[htbp]
\centerline{
\includegraphics[width=.5\linewidth]{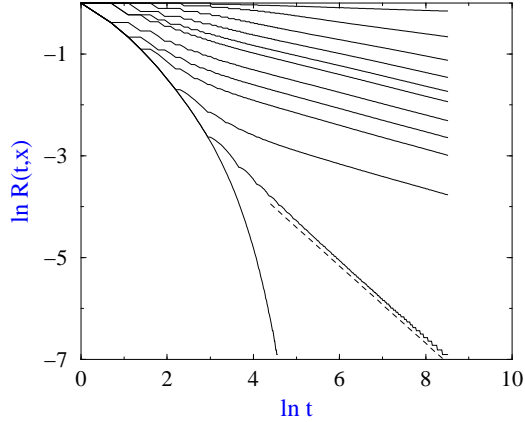}
}
\caption{$R(t,x)$ at $T=0.75\,T_c$ (corresponding to $m_{\rm{eq}}\approx 0.9$)
for the three-dimensional Ising model.
From bottom to top:
$x=$ 1, 0.9, 0.8, 0.6, 0.4, 0.2, 0,
$-0.2$, $-0.4$, $-0.6$, $-0.8$, $-0.9$.
The dashed line has slope $-\theta(m_{\rm eq})=-0.26-\frac{1}{2}$.
(The system size is $512^3$.)}
\label{R3d075_cg_bis}
\end{figure}

The case $x=m_{\mathrm{eq}}$ deserves special mention. 
We find 
\[
R(t,m_{\mathrm{eq}})
={\mathcal{P}}(S_{t^{\prime }}/t^{\prime }>m_{\mathrm{eq}%
}\,\text{ for all }t^{\prime }\leq t)\sim t^{-\theta -\frac{1}{2}}, 
\]
which is equivalent to saying that 
\[
\theta (m_{\mathrm{eq}})=\theta +\frac{1}{2} \qquad(0<T<T_c).
\]
The presence of the exponent $\frac{1}{2}$ can be simply understood. 
The decay of $R(t,m_{\mathrm{eq}})$, i.e., of
the probability for the Ising random walker not to cross the line 
$S_{t}=m_{\mathrm{eq}}t$ has two causes. 
On the longer scale it is due to the crossing of domain walls,
on the shorter time scale it is due to thermal fluctuations of $M_{t}$
around $m_{\mathrm{eq}}$. 
These fluctuations are the same as at equilibrium.
For a system at equilibrium the random walk is biased, since in average 
$\left\langle M_{t}\right\rangle =m_{\mathrm{eq}}$. 
The probability for this random walk not to cross the line 
$S_{t}=m_{\mathrm{eq}}t$ decays as $t^{-\frac{1}{2}}$.

Illustrations of this phenomenon are given in figure \ref{r2dlt09_cg} 
for the two-dimen\-sional case, and in figure \ref{R3d075_cg_bis} for 
a three-dimensional cubic lattice.
In these figures $R(t,m_{\eq})$ appears as a separatrix,
with slope $-\theta-\frac{1}{2}$, where 
$\theta\approx 0.22$ in 2D, and $\theta\approx 0.26$ in 3D. 
This last value is in agreement with the measurement of 
refs.~\cite{stauffer,cueille}.
Note that the accuracy is better in 3D than in 2D, though the size used
in the former case is much smaller.
The reason is that, in order to observe the factor $t^{-\frac12}$,
$m_{\eq}$ has to be significantly different from $1$, 
and therefore $T$ close enough to $T_c$.
On the other hand, if $T$ is too close to $T_c$, crossover
effects become important and preclude the observation of
algebraicity of $R(t,x)$.
Due to the temperature dependence of $m_{\eq}$,
a better compromise is found in 3D.
For instance 
$m_{\eq}=0.9$ for $T=0.895\,T_c$ in 2D, 
while this value is already obtained for
$T=0.75\,T_c$ in 3D. ($T_c^{3D}\approx 4.511$ \cite{blote}.)

Note that the temperature chosen, $T=0.75\,T_c$, is above
the roughening transition of the 3D Ising model, $T_R\approx 0.544\,T_c$
\cite{hasenbusch}.
Below $T_R$ the exponents $\theta(x)$, and in particular $\theta $, take
smaller values.

The persistence probability $p_{0}(t)$ is algebraically decaying as 
$t^{-\theta }$ at $T=0$ only.

\section{Discussion}

Let us give a brief summary.
The three temperature regimes, $T>T_c$, $T=T_c$, and $T<T_c$, correspond,
for the process $\sigma _t$, to 
correlations of increasing strength, 
and, correspondingly, to different behaviours of thesum $S_t$.
\begin{itemize}
\item At high temperature, for $1\ll s\sim t$,
the two-time autocorrelation function
$C(t,s)=\left\langle \sigma _{s}\sigma _{t}\right\rangle$ 
($s<t$) is short ranged and stationary:
$$
C(t,s)\sim\e^{-(t-s)/t_{\eq}},
$$ 
hence $S_t$ obeys the central limit theorem, and
\[
(S_{t})_{\typ}\sim t^{\frac{1}{2}}.
\]

\item At criticality, for $1\ll s\sim t$,
$$
C(t,s)=s^{-2\beta /\nu z_{c}}\,g_{c}\left( \frac{s}{t}\right) ,
$$
implying
$$
(S_{t})_{\mathrm{typ}}\sim t^{1-\beta /\nu z_{c}}. 
$$

\item In the low-temperature phase, for $1\ll s\sim t$, 
$$
C(t,s)= m_{\mathrm{eq}}^{2}\,(T)\,g\left( \frac{s}{t}\right) .
$$
implying
$$
(S_{t})_{\mathrm{typ}}\sim t. 
$$
\end{itemize}

In the first two cases ($T\geq T_c$), the probability density of 
$S_t/t\equiv M_t$ is peaking,
while that of the scaling variable $S_t/(S_{t})_{\mathrm{typ}}$ is
converging to a Gaussian, for $T>T_c$, and to the function $\varphi$
(cf eq. (\ref{fM_crit})), 
for $T=T_c$.

Below $T_{c}$, the probability density $f_{M}(t,x)$ concentrates on 
$[-m_{\mathrm{eq}},m_{\mathrm{eq}}]$, as $t\to\infty$.
In the variable $x/m_{\mathrm{eq}}$, 
it can be rescaled onto a universal curve, which is the 
$T=0$ limiting distribution $f_M$,
singular at $x\rightarrow \pm 1$, with singularity exponent equal to 
$\theta -1$ 
(see figure~\ref{dist0_cg}).
Another alternative definition of the persistence exponent at finite
temperature is provided by the decay of $R(t,m_{\eq})\sim t^{-\theta-\frac12}$.

In summary, the central limit theorem for $S_t$ is violated for any $T\leq T_c$.
On the other hand ergodicity is broken for $T<T_c$, since $M_t$, 
the temporal mean of the spin $\sigma_t$, remains distributed, as $t\to\infty$,
instead of converging to the average $\langle \sigma_t\rangle$.

We conclude by a last comment.
The central limit theorem can be violated in essentially two ways. 
Either by adding identically distributed random variables, 
all of the same orderof magnitude, i.e. with a narrow common distribution,
but otherwise strongly correlated. 
Or by adding independent identically distributed random variables,
with a broad common distribution.
The violation of the central limit theorem for $S_t$ at $T=0$
can be viewed in either of these two ways.
On one hand, by its very definition,  $S_t$ is the sum of the correlated
random variables $\sigma_t$, as discussed in this paper.
On the other hand, 
$S_t$ can also be viewed as the alternating sum of the intervals of time 
$\tau_1$, $\tau_2\ldots$, 
between two sign changes of $\sigma_t$, which are broadly distributed.
Though these intervals of time are not independent, 
the violation of the central limit
theorem is nevertheless essentially due to the algebraic tail 
$\sim \tau ^{-(1+\theta )}$ of their distribution.
The case $0<T\leq T_c$ is more difficult to interpret this way because of
the occurrence of thermal flips.
A natural model to consider however, is one where the intervals of time $\tau_i$
are independent with distribution $\rho(\tau)\sim\tau^{-1-\theta}$ \cite{baldass}.
Though this is a simplification of reality, it leads nevertheless to
the same classification of behaviours for $C(t,s)$ and $(S_t)_{\typ}$
as summarized above,
with respectively $\theta>2$ corresponding to $T>T_c$, $1<\theta<2$ 
to $T=T_c$ and $\theta<1$ to $T<T_c$ \cite{glRenew}.

\bigskip\noindent{\bf Acknowledgments.}
We thank J.P. Bouchaud and J.M. Luck for fruitful discussions.




\end{document}